\documentclass[12p,preprint, showkeys]{revtex4-1}
\usepackage{amsmath,amssymb,graphicx,bm}
\usepackage{epsfig,subfigure}
\usepackage[utf8]{inputenc}

\begin{document}
\author{Yamen hamdouni}

\title{Effect of coupling asymmetry on the spontaneous decay and entanglement of two atoms in a common cavity}
\affiliation{Physics department, University Mentouri, Constantine, Algeria}

\begin{abstract}
We investigate the dynamics and the spontaneous decay of two identical  two-level atoms  into the vacuum in a cavity at low temperature.  We analytically derive the  survival probability and the amplitudes of the state  of the system by calculating the  exact  self-energy when the atoms are both initially excited. It is found that moderate asymmetry enhances the decay of the survival probability and increases   the entanglement production in  either lossless and lossy cavities. 
\end{abstract}

\maketitle

\section{Introduction}
The spontaneous decay of atoms interacting with external fields is a central subject in cavity QED that attracted a great deal of attention. The theoretical framework which enabled the study of the coupling of atoms to electromagnetic fields is the basis of  the domain of quantum optics \cite{knight1,loudon,aspect,scully}. In particular, the model of the two-level atom has been the primary tool used to deal with such kind of problems \cite{frasca}. The most successful model is undoubtedly  the Jaynes-Cummings model which consists of a two-level atom coupled to a single-mode bosonic field \cite{larson}. It has been used in different contexts in particular in  the field of quantum information. Many generalizations have been proposed and  investigated by several authors \cite{knight2,roa,zhang,yang}.

In this work, we propose to study the coupling of two two-level atoms in common cavity to the electric field at low temperature when both atoms are initially excited. Our primary goal is investigate the effect of the asymmetry in the coupling of the atoms on their dynamics. The emphasis is  on the evolution of the survival probability for both strong an weak coupling regimes.

In section \ref{sec2} we introduce the model, along with the initial states of the system under investigation. In section \ref{sec3}, we solve the Schrödinger equation, and we derive the self-energy and compute  the different amplitudes and populations. Section \ref{sec4} is devoted to the  case of lossless cavity, where we investigate the evolution of the populations  and the entanglement. In section \ref{sec5} we deal with  the case of a lossy cavity where the atoms exhibit spontaneous decay. We calculate the spectral functions and we study the decay of the survival probability. The weak-coupling regime is dealt with in section \ref{sec6} where we derive the Markovian master equation to second order, and we investigate the entanglement evolution in this regime. We end the paper with a brief conclusion.
\section{Model\label{sec2}}
We consider two  identical two-level atoms interacting with an electromagnetic field  in a  cavity at zero temperature. The energy splitting of  each atom  will be designated  by $\hbar \omega_0 $, which refers to  the difference between the energy of the  excited state $|e\rangle_i$ and the energy of the ground state $|g\rangle_i$ ($i=1,2$).  The full Hamiltonian of the   system can be written as
\begin{equation}
 H=H_0+H_I
\end{equation}
where $H_0$ is  the free Hamiltonian and $H_I$ describes the coupling between the atoms and the field in the cavity. The former reads
\begin{equation}
 H_0=\hbar\omega_0 \sigma_1^+\sigma_1^-+\hbar\omega_0 \sigma_2^+\sigma_2^-+\sum_{\vec{k}}\hbar\omega_k a^\dag_{\vec k} a_{\vec k}
\end{equation}
with  $\sigma_i^\pm$ being the usual Pauli matrices, which can be expresses in terms of the exited and the ground states as (we drop the index):
\begin{equation}
 \sigma^+=|e\rangle\langle g|, \qquad \sigma^-=|g\rangle\langle e|.
\end{equation}
The operators  $a^\dag_{\vec k}$ and $ a_{\vec k}$, on the other hand, are the creation and annihilation operators associated with  mode ${\vec k}$ of the cavity whose energy is $\hbar\omega_k$ .

We shall adopt  the dipole-moment   approximation for the interaction Hamiltonian, namely:
\begin{equation}
 H_I=-\vec{d_1}.\vec{E}-\vec{d_2}.\vec{E}
\end{equation}
where $\vec{d_i}$ is the dipole moment operator of atom $i$. The latter approximation is a plausible one that is widely used in the field of quantum optics and has proven to be of significant relevance in cavity QED. From this perspective, the quantized  electric field reads
\begin{equation}
 \vec{E}=\sum_{\vec{k}}\sqrt{\frac{\hbar\omega_{\vec{k}}}{2\epsilon_0V}}(a_{\vec{k}} e^{i\vec{k} \vec{r}}e^{-i\omega_k t}+a^\dag_{\vec{k}} e^{-i\vec{k} \vec{r}}e^{i\omega_k t})\vec{\epsilon}_{k}
\end{equation}
with $V$ being  the volume of the cavity and $\vec{\epsilon}_{k}$ designates the polarization vector of mode $\vec{k}$.

Within the  rotating-wave approximations, the interaction Hamiltonian simplifies to
\begin{equation}
 H_I=\sum_{\vec{k}} \left(g_{1k}\sigma_1^+a_{\vec{k}}+g^{*}_{1k}\sigma_1^-a^\dag_{\vec{k}}\right)+ \left(g_{2k}\sigma_1^+a_{2\vec{k}}+g^{*}_{2k}\sigma_1^-a^\dag_{\vec{k}}\right)
\end{equation}
where $g_{ik}$ is the coupling constant of atom number $i$  to mode $\vec{k}$  of the field. In particular, for a linearly polarized field along the $z$ direction,  the coupling constant  is shown to be
\begin{equation}
 g_{ik}=\langle e |d_i|g\rangle \sqrt{\frac{\hbar\omega_{k}}{2\epsilon_0V}} \sin{k z}
\end{equation}

 We assume that the atoms are initially excited and uncorrelated with  the field. Since  the cavity is at zero temperature,  the latter is effectively represented by a  multimode bosonic  field  in its vacuum state. Therefore, the initial state of the full system is given by
 \begin{equation}
  |\phi(0)\rangle=|ee\rangle \otimes |vac\rangle\equiv |ee, vac\rangle, \qquad |vac\rangle =\bigotimes_{\vec k} |0_{\vec k}\rangle
 \end{equation}
 Here we have introduced the state  $|ee\rangle=|e\rangle_1\otimes |e\rangle_2$ along with the Fock  states  $|n_{\vec k}\rangle$  which refer to those states  containing precisely $n$ excitations in mode $\vec{k}$, while the other modes contain zero excitations.
It is straightforward to show that the model Hamiltonian conserves the total number of excitations. Therefore,  the state of the system evolves in the time according to
\begin{widetext}
 \begin{eqnarray}
 |\phi(t)\rangle&=&e^{-2i \omega_0 t}\psi_0(t)|ee, vac\rangle+\sum\limits_{\vec k} e^{-i(\omega_0+\omega_{\vec{k}}) t}\left[\psi_{ k}^1(t)|eg,1_{\vec k}\rangle  +\psi_{ k}^2(t)|ge,1_{\vec k}\rangle \right]\nonumber \\&+&\sum\limits_{\vec k}e^{-2i \omega_k t}\chi_{ k}(t)|gg,2_{\vec k}\rangle
\end{eqnarray}
\end{widetext}
Evidently, the functions $ \psi_{ k}^1(t)$ and $ \psi_{ k}^2(t)$ describe the probability amplitudes corresponding to the processes in which  at most one of the atoms remains excited while the other one decays into the cavity field. On the other hand, the  amplitudes  $\chi_{ k}(t)$ correspond to the cases where both atoms  simultaneously  occupy the  ground states. In the next sections, we  deal essentially with the evolution in time of the latter amplitudes. 
\section{Time evolution \label{sec3}}
There exist several techniques that allow for the derivation of the exact evolution of the wave function $|\phi(t)\rangle$ such as the resolvent operator method. However, it turns out that the problem may be approached directly by solving the Schrödinger equation:
\begin{equation}
 H |\phi(t)\rangle =i \hbar \frac{d|\phi(t)\rangle}{dt}
\end{equation}
To this end, let us introduce the Laplace transform
\begin{equation}
 \tilde f(z)=\int\limits_0^\infty f(t) e^{-z t} dt
\end{equation}
and the detuning
$$ \Delta_k=\omega_0-\omega_k$$
Notice  that in order  to make the notation clear,   any quantity with a tilde, whose argument is denoted by $z$, refers to the Laplace transform. 

By direct substitution of the wave function into the Schrödinger equation, and by Laplace transforming the resulting differential equations for the amplitudes, it follows that (from here on we set  $\hbar=1$)
\begin{widetext}
 \begin{eqnarray}
 z \tilde\psi_0 (z)-1 &=&-i\sum\limits_{{\vec k}}g_{2k} \tilde\psi_{ k}^1(z-i\Delta_k)+g_{1k}\tilde \psi_{ k}^2(z-i\Delta_k)\\
 z  \tilde\psi_{ k}^1(z)&=&-i [ g^{*}_{2k} \tilde \psi_{0}(z+i\Delta_k)+\sqrt{2}g_{1k}\tilde \chi_{ k}(z-i\Delta_k)]\\
  z  \tilde\psi_{ k}^2(z)&=&-i [ g^{*}_{1k} \tilde\psi_{0}(z+i\Delta_k)+\sqrt{2} g_{2k}\tilde \chi_{ k}(z-i\Delta_k)]\\
  z\tilde\chi_{ k}(z)&=&-i\sqrt{2}[ g^{*}_{1k} \tilde\psi_{ k}^1(z+i\Delta_k)+ g^{*}_{2k}\tilde\psi_{ k}^2(z+i\Delta_k)]
\end{eqnarray}
\end{widetext}
\ 
\\
\\
 The relative contribution  of the  amplitudes $ \tilde\psi_{ k}^1(z)$ and $ \tilde\psi_{ k}^2(z)$ depends on the spatial position of the atoms within the cavity. It can be shown that they are related to each other by 
\begin{equation}
 \frac{ \tilde\psi_{ k}^2(z)}{ \tilde\psi_{ k}^1(z)}=\eta_k^{-1} h(z)
\end{equation}
where  $$\eta_k=\frac{g^{*}_{2k}}{g^{*}_{1k}}$$ and 
\begin{equation}h(z)=\frac{z(z-i\Delta_k)+2(|g_{1k}|^2-|g_{2k}|^2)}{z(z-i\Delta_k)+2(|g_{2k}|^2-|g_{1k}|^2)}
 \end{equation}
 
 In order to study the probability of spontaneous decay of the atoms, we  solve for $ \psi_0 (z)$ to find
\begin{equation}
 \tilde\psi_0 (z)=\frac{1}{z+\Sigma(z)}
\end{equation}
where the self-energy function is given by
\begin{widetext}
\begin{equation}
 \Sigma(z)=
\sum\limits_{{\vec k}}\frac{|g_{1k}|^2(\eta_k^2+ h(z-i\Delta_k))(z-2i\Delta_k)}{[(z-i\Delta_k)(z-2i\Delta_k)+2|g_{1k}|^2(1+ h(z-i\Delta_k))]}
\end{equation}
\end{widetext}
It follows that the time dependence of the wave function $\psi_0$ takes the form 
\begin{equation}
 \psi_0(t)=-\frac{e^{2i\omega_0 t}}{2\pi}\int\limits_{-\infty+i c}^{+\infty+i c}\frac{e^{-i E t} dE}{E-2\omega_0-\Sigma_{\rm ret}(E)} \label{main1}
\end{equation}
where $c$ is  real, and  the retarded self energy is given by
\begin{equation}
 \Sigma_{\rm ret}(E)=-i\Sigma(-i(E+i0^+-2\omega_0))
\end{equation}
Similarly, we obtain that
\begin{widetext}
\begin{equation}
 \psi^1_k(t)=-\frac{e^{i(\omega_0+\omega_k) t}}{2i\pi}\int\limits_{-\infty+i c}^{+\infty+i c}g^{*}_{2k} \frac{e^{-i E t} \tilde\psi_0(-i(E-2\omega_0)) dE}{E-\omega_0-\omega_k-\frac{2|g_{1k}|^2(1+ h(-i(E-\omega_0-\omega_k))}{E-2\omega_k}}
\end{equation}
\end{widetext}
and 
\begin{widetext}
\begin{equation}
 \psi^2_k(t)=-\frac{e^{i(\omega_0+\omega_k) t}}{2i\pi}\int\limits_{-\infty+i c}^{+\infty+i c}g^{*}_{1k} \frac{e^{-i E t} \tilde\psi_0(-i(E-2\omega_0)) h(-i(E-\omega_0-\omega_k)) dE}{E-\omega_0-\omega_k-\frac{2|g_{2k}|^2(1+ h(-i(E-\omega_0-\omega_k))}{E-2\omega_k}}
\end{equation}
\end{widetext}

The survival probability  is thus expressed as
\begin{equation}
 P_{ee}(t)=|\psi_0(t)|^2
\end{equation}
which is the probability that both atoms remain excited. Furthermore, the populations of the  states $|eg\rangle$  and $|ge\rangle$ are given  respectively by
\begin{equation}
 P_{eg}(t)=\sum\limits_{\vec k}|\psi_k^1|^2
\end{equation}
 and 
\begin{equation}
 P_{ge}(t)= \sum\limits_{\vec k}|\psi_k^2|^2.
 \end{equation}
 The probability that both atoms  occupy simultaneously their ground states is  thus equal to 
 \begin{equation}
 P_{gg}(t)=\sum\limits_{\vec k}|\chi_k|^2=1-P_{ee}(t)-P_{eg}(t)-P_{ge}(t).
 \end{equation}
\section{Case of lossless cavity  \label{sec4}}
Let us suppose that the cavity is  lossless and that the field consists of a single mode of frequency $\omega_f$. The sum in the self energy reduces to a single term and the denominator in equation (\ref{main1}) possesses four real poles $E_j$.
 Hence, by transforming the integral in equation (\ref{main1}) into closed contour integral round a semi-circle in the lower complex half-plane   it follows that

\begin{eqnarray}
 \psi_0(t)=i e^{i2\omega_0 t} \sum_{j} \frac{e^{-iE_j t}}{1-\frac{d\Sigma_{\rm ret}(E)}{dE}|_{E=E_j,\omega=\omega_f}}
\end{eqnarray}
Following the same procedure, it can be shown that
\begin{eqnarray}
 \psi^1_f(t)&=&- \eta g \sum_{j}  e^{i(\omega_0+\omega_f)t}  \frac{e^{-iE_j t}}{\frac{d F_1(E)}{dE}|_{E=E_j,\omega_k=\omega_f}}
\end{eqnarray}
where 
\begin{equation}
 F_1(E)= \frac{E-\omega_0-\omega_k-\frac{2g^2(1+ h(-i(E-\omega_0-\omega_k))}{E-2\omega_k}}{\tilde\psi_0(-i(E-2\omega_0))}
\end{equation}
and we have used the notations  $g=g_1$ and $\eta=g_2/g_1$. The other amplitudes $\psi^2_f(t)$ and $\chi_f(t)$ may be calculated with the same method.
 By taking the partial trace over the cavity mode, it is easy to show that the reduced density matrix of the two-atom system takes the form
 \begin{equation}
  \rho(t)=\begin{pmatrix}
           P_{ee} &&  0 && 0 && 0 \\
           0 &&  P_{eg}  &&\psi^{1}_f(t)  \psi^{2*}_f(t)  &&0 \\
           0 &&   \psi^{1*}_f(t) \psi^{2}_f(t)  && P_{ge}  &&0\\
           0&& 0 &&0&&  & P_{gg}
          \end{pmatrix}
 \end{equation}
\begin{figure}[htb!]
 \centering
 \includegraphics{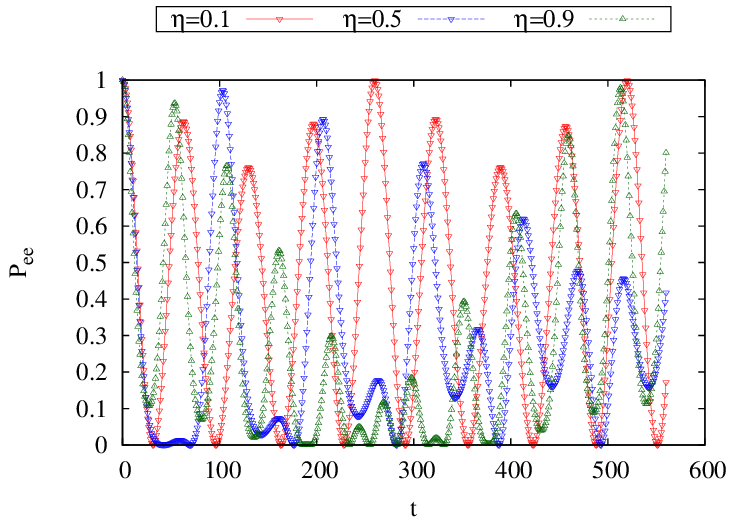}
 \caption{The evolution in time of the population $P_{ee}$ with $\omega_0=\omega_f=7$, $g=0.5$ for different values of the asymmetry parameter $\eta$}\label{fig1}
\end{figure}
\begin{figure}[htb!]
 \centering
 \includegraphics{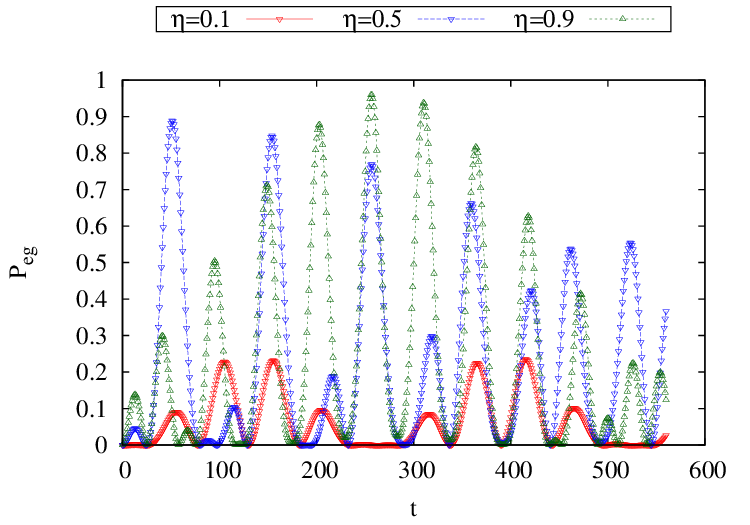}
 \caption{The  time evolution  of the Population $P_{eg}$ with $\omega_0=\omega_f=7$, $g=0.5$ for different values of the asymmetry parameter $\eta$}\label{fig2}
\end{figure}

\begin{figure}[htb!]
 \centering
 \includegraphics{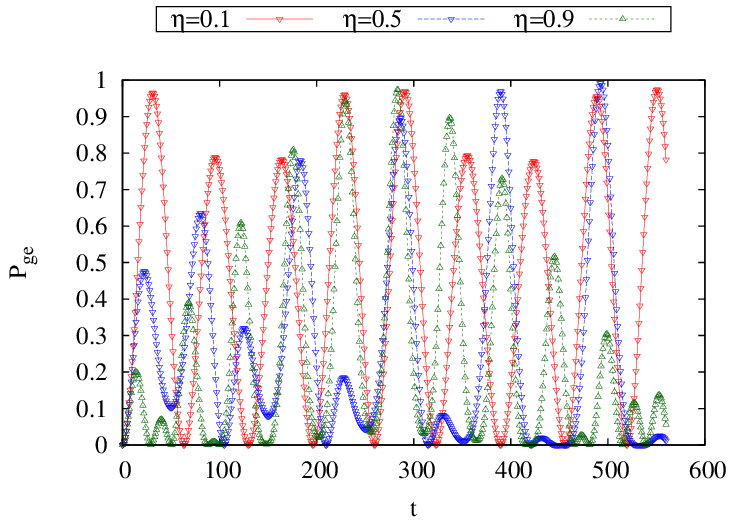}
 \caption{The evolution in time of the Population $P_{ge}$ with $\omega_0=\omega_f=7$, $g=0.5$ for different values of the asymmetry parameter $\eta$}\label{fig3}
\end{figure}
\begin{figure}[htb!]
 \centering
 \includegraphics{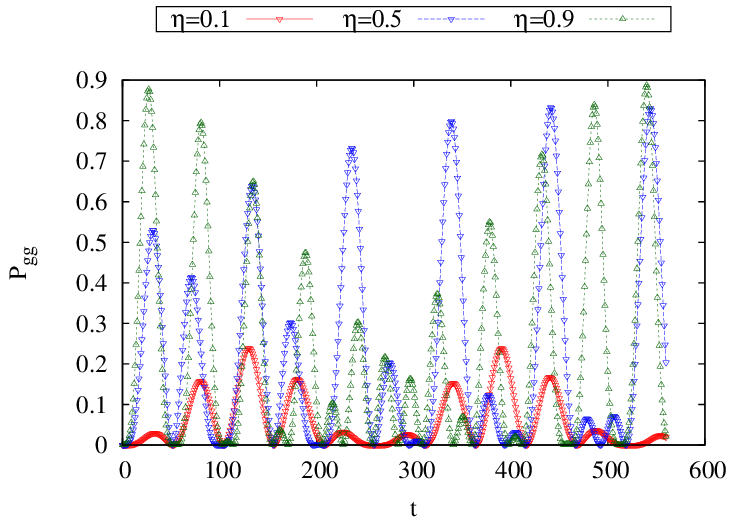}
 \caption{The evolution in time of the Population $P_{gg}$ with $\omega_0=\omega_f=7$, $g=0.5$ for different values of the asymmetry parameter $\eta$}\label{fig4}
\end{figure}
In figure \ref{fig1} we display the variation in time of  the survival probability for different values of the parameter $\eta$. We see that it does not follow a regular pattern as regard of the dependence of the  magnitude of the survival probability on the latter parameter in the course of the time. This translates the back-action of the cavity on either atom. Indeed, at  short times the decay is as expected faster when the coupling is nearly symmetric. At later times the revival of the survival probability is much important for either strong or weak asymmetry. Surprisingly, intermediate values of $\eta$ yield the least revival of the population compared to the strong asymmetry (weak ratio $\eta$). On the contrary, at much longer time, the  survival probability  corresponding to the nearly symmetric  coupling is more likely to decrease, while the case of strong asymmetry  yields the largest survival probability.   
The population of the states $|eg\rangle$ and $|ge\rangle$ is illustrated  in figures \ref{fig2} and \ref{fig3}. We notice that  for $\eta=0.5$, the second atom, even though weakly coupled to the cavity, is much more likely to make transition to its ground state at intermediate times  compared to that of the first atom whose coupling is larger. In both figures, the nearly symmetric coupling leads to the smallest population of these states. This is due to two-photon processes inherent for the on-resonance  dynamics of the atoms as seen in figure \ref{fig4}. Nevertheless, the probability of both atoms being in their ground state is larger for the moderate value of of $\eta=0.5$ where  one-photon absorption is dominant in the zero-detuning case.

\begin{figure}[htb!]
 \centering
 \includegraphics{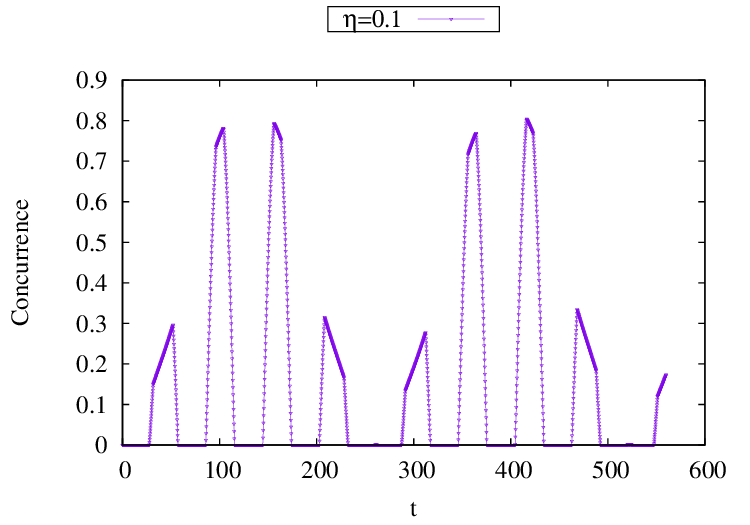}
 \caption{Concurrence $C(\rho)$ as a function of time  for $\omega_0=\omega_f=7$, $g=0.5$ and $\eta=0.1$}\label{fig5}
\end{figure}
\begin{figure}[htb!]
 \centering
 \includegraphics{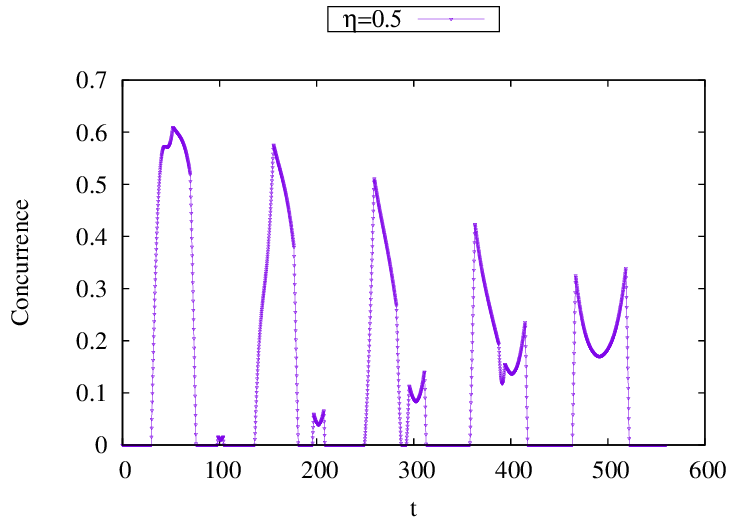}
 \caption{Concurrence $C(\rho)$ as a function of time  for $\omega_0=\omega_f=7$, $g=0.5$ and $\eta=0.5$ }\label{fig6}
\end{figure}
\begin{figure}[htb!]
 \centering
 \includegraphics{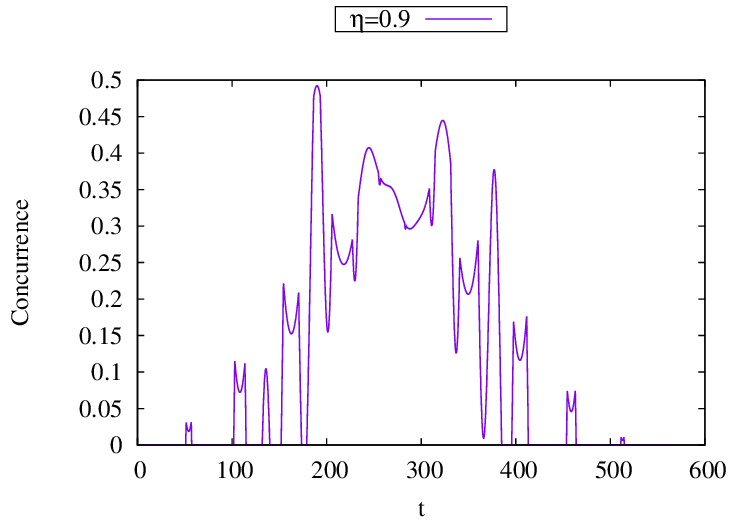}
 \caption{Concurrence $C(\rho)$ as a function of time  for $\omega_0=\omega_f=7$, $g=0.5$ and $\eta=0.9$}\label{fig7}
\end{figure}
Let us now investigate the entangling power of the cavity. As is known there could be no entanglement generation in the strictly symmetric case $\eta=1$. To check whether the state of the two atoms, initially separable, becomes entangled,   we use the   concurrence $C(\rho)$  as a measure of entanglement \cite{wootters,vidal}. The concurrence is defined by
\begin{equation}
 C(\rho)={\rm max}\{0, \sqrt{\lambda_1}- \sqrt{\lambda_2}- \sqrt{\lambda_3}- \sqrt{\lambda_4}\},
\end{equation}
where $ \lambda_1 > \lambda_2> \lambda_3 >\lambda_4$ denote  the eigenvalues of the matrix
\begin{equation}
\tilde\rho(t) = \rho(t)(\sigma_y \otimes \sigma_y )\rho^*(t)(\sigma_y \otimes \sigma_y ).
\end{equation}
The values of the concurrence range from zero for separable states to one for maximally entangled states.

From figures \ref{fig5}, \ref{fig6} and \ref{fig7} we clearly notice that the atoms get more entangled as the coupling asymmetry increases. The generation of entanglement is more pronounced  for smaller values of the parameter  $\eta$. Therefore, strong atom-field interaction does not yield the largest entanglement  of the two-atom system in all cases. What matters in this multimode Jaynes-Cummings model is rather the complex effects caused by the asymmetric coupling to the cavity.     For such small values, we can identify periods of the time where the concurrence  increases or decreases almost linearly as a function of time. This is a nontrivial behavior that indicates a certain regularity in the dynamics,   which can also be seen for relatively intermediate values of the parameter $\eta$. The behavior of the concurrence is much smoother in figure \ref{fig7} corresponding to weaker asymmetry. As   the parameter $\eta$ gets closer to unity, the concurrence decreases to completely vanish at $\eta=1$. This turns out to be also the case as $\eta$ decreases to zero. Hence the regime of intermediate values of the asymmetry parameter is the more interesting one as long as entanglement is concerned. 
\section{Atoms in lossy cavity\label{sec5}}
Real cavities are not perfectly isolated from the outer environment. To account for this fact, we use the more realistic model that consists of the two atoms being in a  lossy cavity where the field is rather characterized by a spectral density $J(\omega)$. By doing so, it it is possible to encompass spontaneous decay of the atoms into the cavity. For our purpose, we shall assume that the spectral density  is a  Lorentzian distribution, namely:
\begin{equation}
 J(\omega)=\frac{g^2}{\pi}\frac{\Gamma}{(\omega-\omega_f)^2+\Gamma^2}
\end{equation}
 where  $\omega_f$ designates the cavity frequency and  $\Gamma$ is the width of the spectrum. Furthermore   $g$  is the bare  effective coupling of  the model. 
  It follows that the retarded self energy  takes the form
 \begin{eqnarray}
  \Sigma_{\rm ret}(E)&=&-\frac{2g^2\Gamma}{\pi}\left(\frac{\eta^2}{1+\eta^2}\right) \oint_C\frac{(E-2z)dz}{((z-\omega_f)^2+\Gamma^2)(z-\omega_1(E))(z-\omega_2(E))}\nonumber \\
  &-&\frac{g^2\Gamma}{\pi}\left(\frac{(1-\eta^2)^2}{1+\eta^2}\right) \oint_C\frac{dz}{((z-\omega_f)^2+\Gamma^2)(z-E+\omega_0)}
 \end{eqnarray}
 where
\begin{eqnarray}
 \tilde\Omega(E)&=&\sqrt{(E-2\omega_0)^2+16g^2(1+\eta)^2}\\
 \omega_1(E)&=&\frac{1}{4}(3E-2\omega_0-\tilde\Omega(E))\\
 \omega_2(E)&=&\frac{1}{4}(3E-2\omega_0+\tilde\Omega(E))
\end{eqnarray}
We choose the  contour of integration to be a semicircle  closed in the upper half-plane. Afterwards, using the residue theorem we  find that the imaginary part of the retarded self energy is 
\begin{eqnarray}
 {\rm Im}\Sigma_{\rm ret}(E)&=&-\frac{8g^2\Gamma}{\tilde\Omega(E)}\left(\frac{\eta^2}{1+\eta^2}\right)\Bigl[\frac{E-2\omega_1(E)}{(\omega_1(E)-\omega_f)^2+\Gamma^2}-\frac{E-2\omega_2(E)}{(\omega_2(E)-\omega_f)^2+\Gamma^2}\Bigr]\nonumber \\
 &-& \left(\frac{(1-\eta^2)^2}{1+\eta^2}\right)\frac{g^2\Gamma}{(\omega_f+\omega-E)^2+\Gamma^2}
\end{eqnarray}
On the other hand, it can be shown that the real part is
\begin{eqnarray}
 &&{\rm Re}\Sigma_{\rm ret}(E)=-\frac{8g^2\Gamma}{\tilde\Omega(E)}\left(\frac{\eta^2}{1+\eta^2}\right)\Biggl[\frac{(\omega_1(E)-\omega_f)(2\omega_f-E)-2\Gamma^2}{(\omega_1(E)-\omega_f)^2+\Gamma^2}\nonumber \\   &&  \hspace*{3cm} -\frac{(\omega_2(E)-\omega_f)(2\omega_f-E)-2\Gamma^2}{(\omega_2(E)-\omega_f)^2+\Gamma^2}\Biggr]\nonumber \\
 && \hspace*{5cm} -\left(\frac{(1-\eta^2)^2}{1+\eta^2}\right)\frac{g^2\Gamma(\omega_f+\omega-E)}{(\omega_f+\omega-E)^2+\Gamma^2}
\end{eqnarray}

\begin{figure}[htb!]
 \centering
 \includegraphics{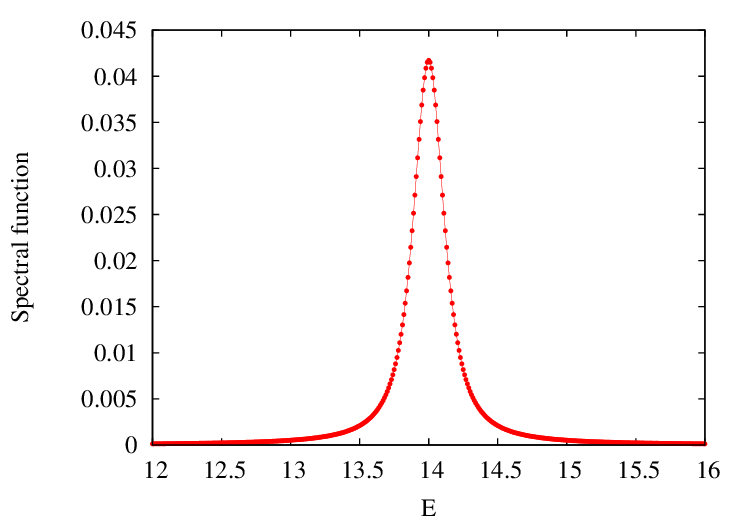}
 \caption{Spectral function $A(E)$  in the weak-coupling regime for $\omega_0=\omega_f=7$, $g=0.05$, $\Gamma=0.1$ and $\eta=0.5.$ }\label{fig8}
\end{figure}
\begin{figure}[htb!]
 \centering
 \includegraphics{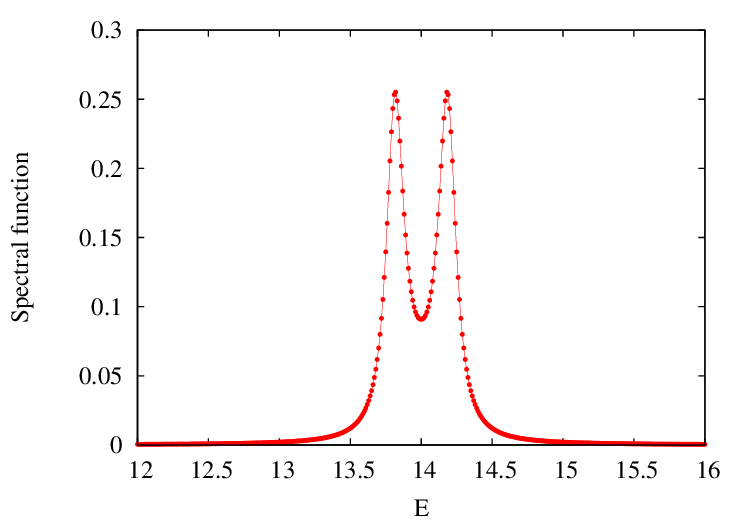}
 \caption{Spectral function $A(E)$  in the strong-coupling regime for $\omega_0=\omega_f=7$, $g=0.1$, $\Gamma=0.05$ and $\eta=0.9.$ }\label{fig9}
\end{figure}
\begin{figure}[htb!]
 \centering
 \includegraphics{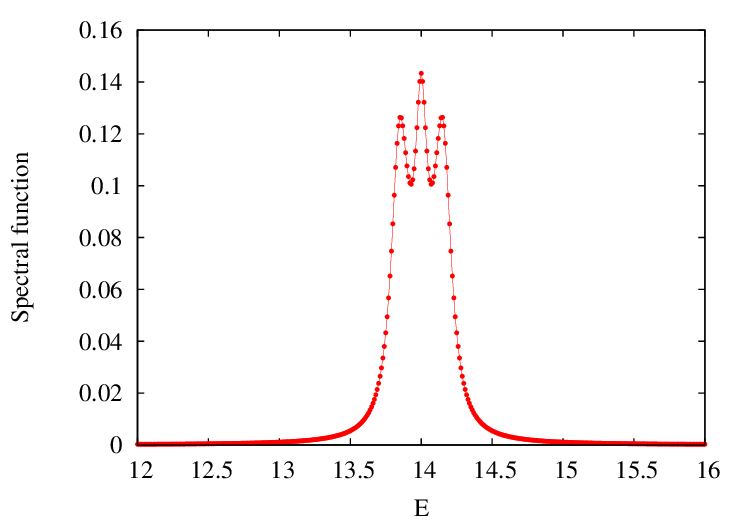}
 \caption{Spectral function $A(E)$  in the strong-coupling regime for $\omega_0=\omega_f=7$, $g=0.1$, $\Gamma=0.05$ and $\eta=0.5$. }\label{fig10}
\end{figure}

We shall write the retarded self-energy as 
$$\Sigma_{\rm ret}(E)={\rm Re}\Sigma_{\rm ret}(E)+i {\rm Im}\Sigma_{\rm ret}(E)=\Sigma_1(E)+i \Sigma_2(E)$$
The spectral function of the two atoms is thus given by
\begin{equation}
 A(E)=-\frac{1}{\pi}  \Sigma_2(E)
\end{equation}
We can now distinguish between two regimes of the dynamics, namely the weak-coupling regime corresponding to $g<\Gamma\ll \omega_0$ and the strong-coupling regime associated with $\Gamma <g \ll \omega_0$. In both regimes the rotating-wave approximation holds. The spectral function of the atoms is illustrated in figures \ref{fig8}, \ref{fig9} and \ref{fig10} for both regimes and for different values of the asymmetry parameter $\eta$. We notice that for weak coupling the  spectral function  possesses a single peak in contrast to the strong-coupling regime where for large values of the asymmetry parameter ($\eta=0.9)$ the shape of the function displays two peaks that in fact collapse into a single peak when $\eta$ becomes unity. As we decrease the value of the latter parameter there appear three peaks in the spectral function, which may explain the special features of the dynamics for the moderate values of the asymmetry parameter (see bellow). In all cases the peaks appear close to the energy $E=2\omega_0$ which is the initial energy of the atoms. The exact positions of the maxima depend on the model parameters.

Since $\Sigma(E)$  displays a discontinuity along the real line, it can be shown that
\begin{equation}
 \psi_0(t)=-\frac{1}{\pi}\int\limits_{-\infty}^\infty \frac{e^{-iE t}\Sigma_2(E) dE }{(E-2\omega_0-\Sigma_1(E))^2+\Sigma_2(E)^2}
\end{equation}
\begin{figure}[htb!]
 \centering
 \includegraphics{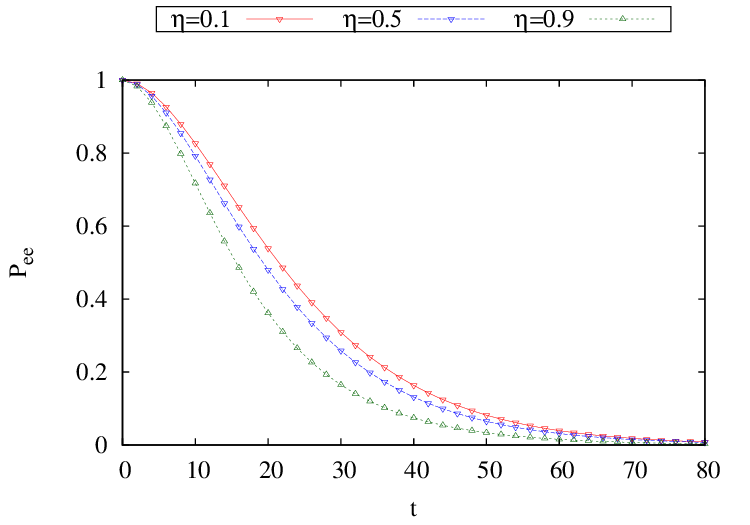}
 \caption{Survival probability in the weak-coupling regime for $\omega_0=\omega_f=7$, $g=0.05$, $\Gamma=0.1$. }\label{fig11}
\end{figure}
\begin{figure}[htb!]
 \centering
 \includegraphics{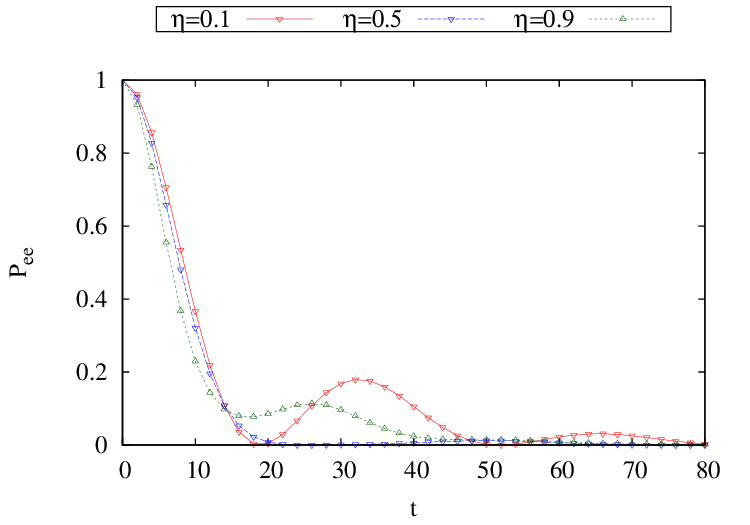}
 \caption{Survival probability in the strong-coupling regime for $\omega_0=\omega_f=7$, $g=0.1$, $\Gamma=0.05$.}\label{fig12}
\end{figure}
\begin{figure}[htb!]
 \centering
 \includegraphics{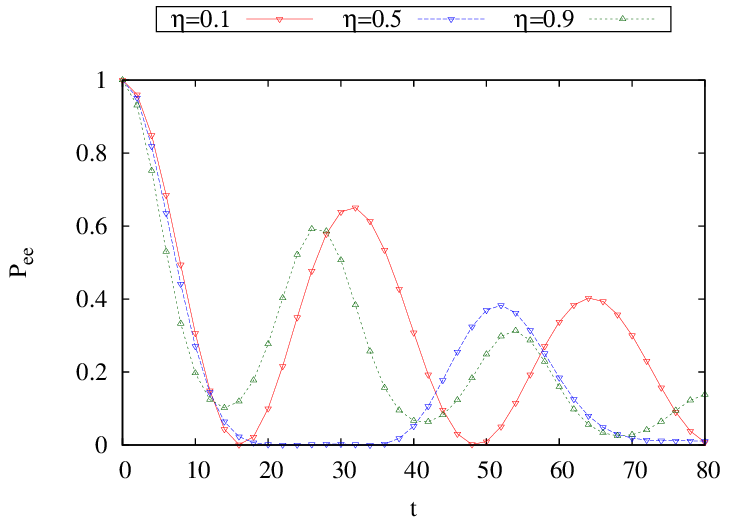}\caption{Survival probability in the strong-coupling regime for $\omega_0=\omega_f=7$, $g=0.5$, $\Gamma=0.05$.}\label{fig13}
\end{figure}
The survival probability for both weak-coupling and strong-coupling is displayed in figures \ref{fig11}, \ref{fig12} and \ref{fig13}. We see that at short time the decay of the survival probability is faster in the case of weak asymmetry. However, in the strong-coupling regime, the revival of the population is larger for strong and weak asymmetry. Moderate values of the parameter $\eta$  yield much noticeable decay.

\section{Weak coupling \label{sec6}}
In the weak coupling limit, under the Born-Markov approximation, the reduced density matrix of the two-atom system satisfies in the interaction picture \cite{petru}
\begin{equation}
 \frac{d\rho}{dt}=-\int\limits_0^\infty {\rm Tr_B}[H(t),[H_I(t-t'),\rho(t-t')\otimes|vac\rangle\langle vac|]] dt'
\end{equation}
To perform the partial trace over the environmental degrees of freedom (denoted here by ${\rm Tr_B}$), we make the transition to the continuum limit. This can be fulfilled by choosing the Lorentzian spectral density such that
\begin{equation}
 \sum_{\vec k} g_{1k} g_{2k}^* \cdots \rightarrow \int\limits_{-\infty}^\infty \eta J(\omega)\cdots =\frac{\eta g^2}{\pi} \int\limits_{-\infty}^\infty d\omega \frac{\Gamma}{(\omega-\omega_f)^2+\Gamma^2} \cdots
\end{equation}
Afterwards, by evaluating the double commutator, it ca be shown that the elements of the reduced density matrix satisfy
\begin{eqnarray}
 \frac{d\rho_{11}}{dt}&=&-(\Gamma_{11}+\Gamma_{22})\rho_{11}(t)\\
 \frac{d\rho_{22}}{dt}&=&\Gamma_{22}\rho_{11}(t)-\Gamma_{11}\rho_{22}(t)+(i\xi_{12}-\Gamma_{12}/2)\rho_{23}(t)-(i\xi_{12}+\Gamma_{12}/2)\rho_{32}(t)\\
 \frac{d\rho_{33}}{dt}&=&\Gamma_{11}\rho_{11}(t)-\Gamma_{22}\rho_{33}(t)-(i\xi_{12}+\Gamma_{12}/2)\rho_{23}(t)+(i\xi_{12}-\Gamma_{12}/2)\rho_{32}(t)\\
 \frac{d\rho_{23}}{dt}&=&\Gamma_{12}\rho_{11}(t)+(i\xi_{12}-\Gamma_{12}/2)\rho_{22}(t)-(i\xi_{12}+\Gamma_{12}/2)\rho_{33}(t)\nonumber \\
 &-&\left(i\xi_{11}-i\xi_{22}+(\Gamma_{11}+\Gamma_{22})/2\right)\rho_{23}(t)\\
 \frac{d\rho_{44}}{dt}&=&\Gamma_{11}\rho_{22}(t)+\Gamma_{22}\rho_{33}(t)+\Gamma_{12}\rho_{23}(t)+\Gamma_{12}\rho_{32}(t)
\end{eqnarray}
where we have introduced the decay rates
\begin{eqnarray}
 \Gamma_{11}=&&\frac{2g^2\Gamma}{{(\omega-\omega_f)^2+\Gamma^2}}, \qquad \Gamma_{22}=\frac{2\eta^2 g^2\Gamma}{{(\omega-\omega_f)^2+\Gamma^2}}\\
 && \qquad \qquad \Gamma_{12}=\frac{2\eta g^2\Gamma}{{(\omega-\omega_f)^2+\Gamma^2}}
\end{eqnarray}
The Lamb shifts on the other hand read as follows:
\begin{eqnarray}
 \xi_{11}=&&\frac{g^2 (\omega-\omega_f)}{{(\omega-\omega_f)^2+\Gamma^2}}, \qquad \xi_{22}=\frac{\eta^2 g^2 (\omega-\omega_f)}{{(\omega-\omega_f)^2+\Gamma^2}}\\
 && \qquad \qquad \xi_{12}=\frac{\eta g^2 (\omega-\omega_f)}{{(\omega-\omega_f)^2+\Gamma^2}}
\end{eqnarray}
\begin{figure}[htb!]
 \centering
 \includegraphics{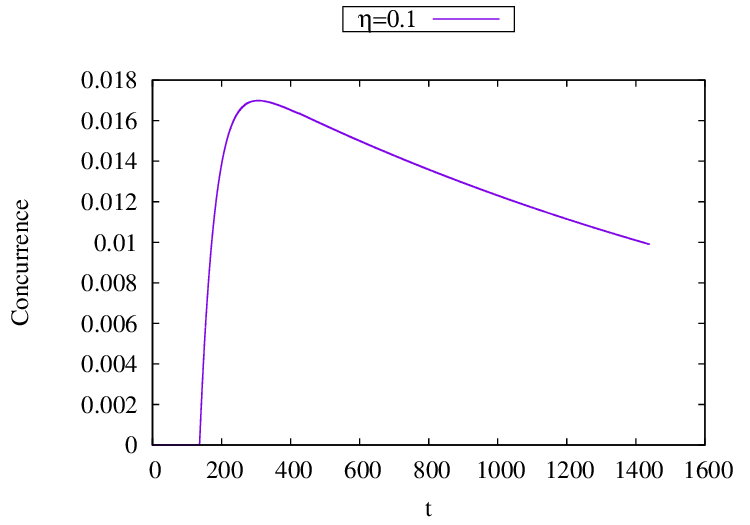}
 \caption{Concurrence  in the weak-coupling regime for $\omega_0=\omega_f=7$, $g=0.05$, $\Gamma=0.1$ and $\eta=0.1$. }\label{fig14}
\end{figure}
\begin{figure}[htb!]
 \centering
 \includegraphics{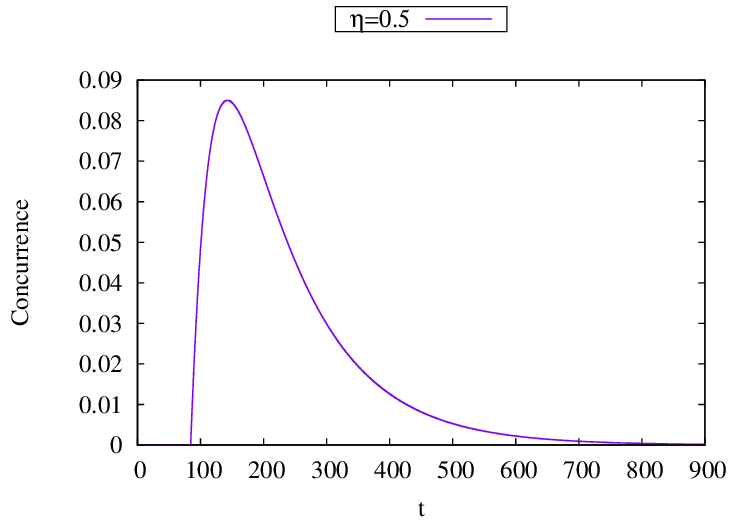}
 \caption{Concurrence  in the weak-coupling regime for $\omega_0=\omega_f=7$, $g=0.05$, $\Gamma=0.1$ and $\eta=0.5$.}\label{fig15}
\end{figure}
\begin{figure}[htb!]
 \centering
 \includegraphics{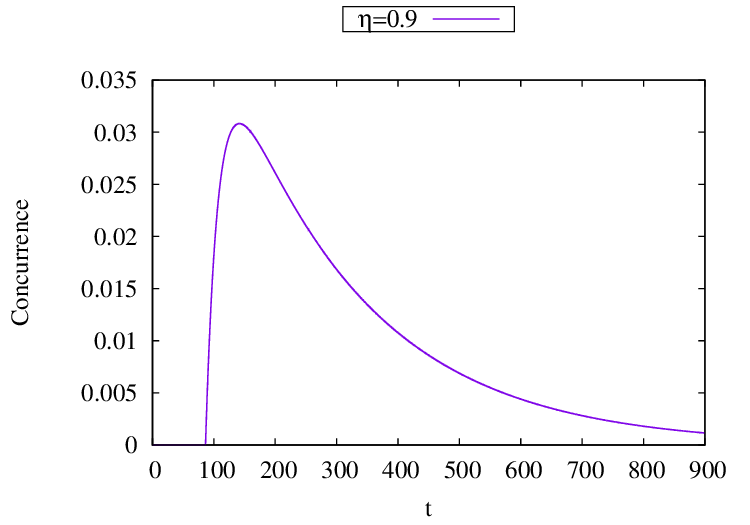}\caption{Concurrence  in the weak-coupling regime for $\omega_0=\omega_f=7$, $g=0.05$, $\Gamma=0.1$ and $\eta=0.9$.}\label{fig16}
\end{figure}
By solving  the above equations, we can evaluate the concurrence which is displayed in figures \ref{fig14}, \ref{fig15} and \ref{fig16} for different values of the parameter $\eta$. The variation is now smooth but  the concurrence is much noticeable for moderate values of the asymmetry. This is in agreement with the case of lossless cavity. The entanglement tend to vanish for either strong and weak asymmetry. Hence even in the Markovian regime, the dynamics of the entanglement preserves the features encountered in the lossless case.
\section{Conclusion}
In conclusion we have studied the influence of the differences in the coupling constants of two two-level atoms coupled to a common field in a cavity at zero temperature. It turns out that moderate asymmetry yield the most pronounced decay of the survival probabilities. It exhibits also larger values of the concurrence, meaning that the cavity has strong entangling power for these values. The entanglement increases as the asymmetry parameter decreases for unity to reach a maximum value at the aforementioned intermediate values, then it tends to decrease as the coupling of one of the atoms gets smaller. These features show the complex interplay between two-photon and single-photon process in the presence of coupling asymmetry. It would be interesting to investigate the effect of the detuning in these cases.

\end{document}